\documentclass[conference]{IEEEtran}

\IEEEoverridecommandlockouts
\usepackage{multirow}%
\usepackage{bbm}%
\usepackage{bm}%
\usepackage{cite}%
\usepackage{amsmath,amssymb,amsfonts,tabularx}%
\usepackage{algorithm, algorithmic}%
\usepackage{graphicx}%
\usepackage{graphics}
\usepackage{textcomp}%
\usepackage{dsfont}%
\usepackage{cleveref}
\usepackage{comment}
\usepackage{gensymb}
\usepackage[long, c3, nocomma]{optidef}
\usepackage{epstopdf}
\usepackage{refcount}
\usepackage{threeparttable}
\usepackage[activate={true,nocompatibility},final,tracking=true,kerning=true,spacing=true,factor=1100,stretch=10,shrink=10]{microtype}
    
\def\BibTeX{{\rm B\kern-.05em{\sc i\kern-.025em b}\kern-.08em
    T\kern-.1667em\lower.7ex\hbox{E}\kern-.125emX}}

\ifCLASSOPTIONcompsoc
    \usepackage[caption=false, font=normalsize, labelfont=sf, textfont=sf]{subfig}
\else
\usepackage[caption=false, font=footnotesize]{subfig}
\fi

\makeatletter

\begin{document}
\title{Hierarchical Multi-timescale Framework for Operation of Dynamic Community Microgrid}

\author{\IEEEauthorblockN{Ashwin Shirsat, Valliappan Muthukaruppan, Rongxing Hu, Ning Lu, Mesut Baran, David Lubkeman, \\ Wenyuan Tang}
\IEEEauthorblockA{Department of Electrical and Computer Engineering\\
North Carolina State University\\
Raleigh, NC 27695, USA\\
\{ashirsa, vmuthuk2, rhu5, nlu2, baran, dllubkem, wtang8\}@ncsu.edu}
\thanks{This material is based upon work supported by the U.S. Department of Energy's Office of Energy Efficiency and Renewable Energy (EERE) under the Solar Energy Technologies Office Award Number DE-EE0008770.}
}

\maketitle
\begin{abstract}
Distribution system integrated community microgrids (CMGs) can restore loads during extended outages. The CMG is challenged with limited resource availability, absence of a robust grid-support, and demand-supply uncertainty. To address these challenges, this paper proposes a three-stage hierarchical multi-timescale framework for scheduling and real-time (RT) dispatch of CMGs. The CMG's ability to dynamically expand its boundary to support the neighboring grid sections is also considered. The first stage solves a stochastic day-ahead (DA) scheduling problem to obtain referral plans for optimal resource rationing. The intermediate near real-time scheduling stage updates the DA schedule closer to the dispatch time, followed by the RT dispatch stage. The proposed methodology is validated via numerical simulations on a modified IEEE 123-bus system, which shows superior performance in terms of RT load supplied under different forecast error cases, outage duration scenarios, and against the traditionally used two-stage approach.
\end{abstract}


The distribution grid resiliency needs to be enhanced to withstand, operate, and recover from disruptions caused by extreme events such as hurricanes and wildfires. Post outage, a resilient grid uses modern automation techniques, algorithms, and information-communication technology to restore loads \cite{restoration_1}. Conventional system restoration strategies use the upstream transmission system along with distribution network reconfiguration post outage. However, such strategies are not effective at times when extreme events disrupt the transmission grid. 
To address this challenge, community microgrids (CMGs) have proved to be very promising \cite{whycmg}. During such extended outages, the CMGs can operate in an islanded manner and ensure continued operation for its local loads. Further, they can also supply some part of the distribution grid by expanding their boundary to accommodate the neighboring nodes. 

In \cite{da_mg_sch_op1}, Yang \textit{et al.} have proposed a two-stage approach for microgrid (MG) scheduling and real-time (RT) dispatch. The scheduling problem is solved for the projected outage duration, and the optimal power flow (OPF) based RT dispatch problem is solved using the scheduling results. In \cite{da_mg_scheduling1}, Gholami \textit{et al.} have proposed a robust optimization-based day-ahead (DA) scheduling for MG resiliency enhancement. In \cite{da_mg_sch_op2}, Qiu \textit{et al.} propose a three-stage formulation for optimal dispatch of MGs for islanded operation under normal conditions. The existing literature covers various aspects of MG energy management. However, a holistic approach for proactive scheduling and dispatch of MGs during emergencies emphasizing uncertainty mitigation, critical load priority, optimal resource allocation for self-sustained operation, and MG support expansion to the neighboring grid has not been clearly addressed. The existing literature's operational objective is on cost minimization, which takes a lower priority during emergencies. 

To address the above limitations, the contributions of our paper are summarized as follows. \textit{First}, we present a three-stage hierarchical multi-timescale (HMTS) model for proactive scheduling and RT dispatch of a CMG having a high penetration of residential behind-the-meter (BTM) PV generators. \textit{Second}, a CMG with dynamic boundary is considered, i.e., the CMG can dynamically expand its boundary to support neighboring system nodes. To incorporate this in the scheduling stage, a modified set of constraints is proposed to optimally decide the CMG boundary expansion. 
\textit{Third}, we focus the objectives of the proposed approach on optimal resource allocation to ensure resource availability at all times, mitigate the impact of forecasting errors, prioritize service to critical loads, and provide reliable grid-forming support.
\vspace{-0cm}
\section{HMTS Problem Formulation}
\begin{figure}[tb]
  \centering
  \includegraphics[width = \linewidth ,keepaspectratio, trim={3.25cm 0cm 2.5cm 0cm},clip]{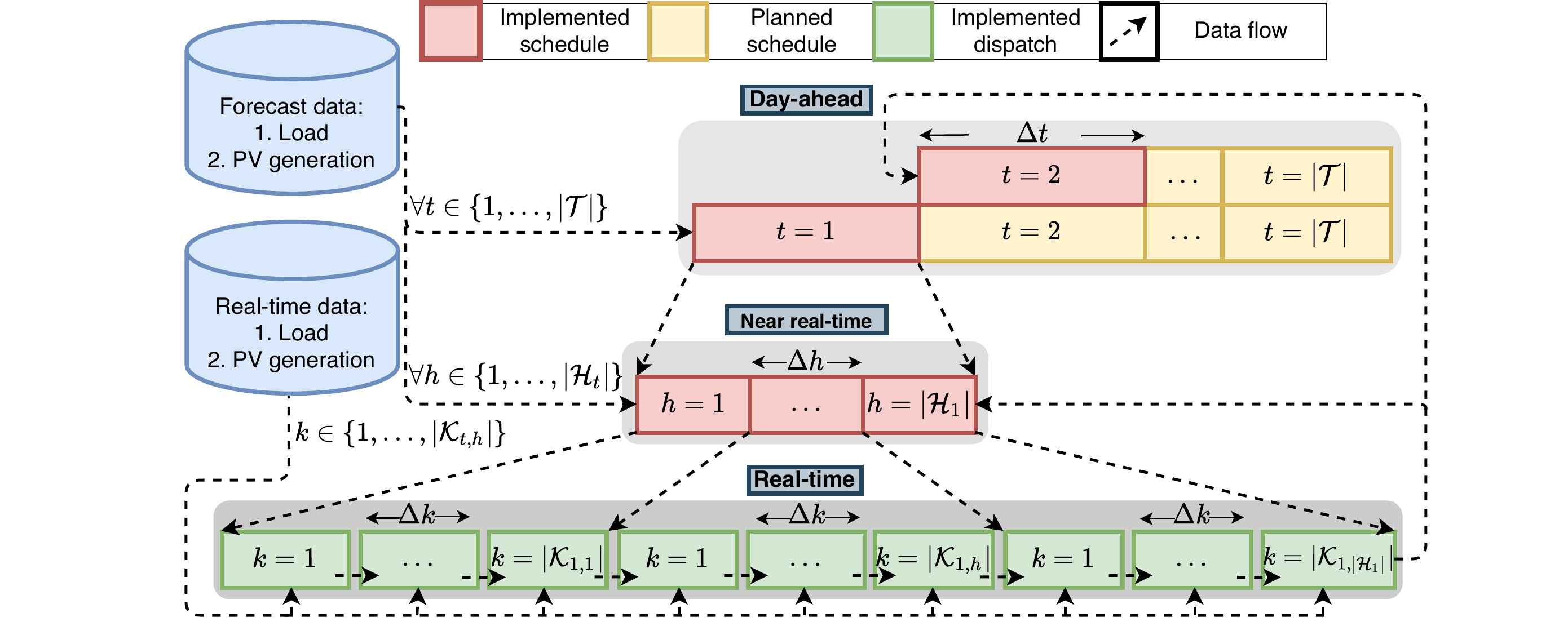}
  \caption{Schematic layout of the proposed HMTS framework.}
  \label{fig:layout}
  \vspace{-0.5cm}
\end{figure}

Minimal load and renewable generation data is available for low-frequency high impact events for making accurate forecasts. Hence, an intermediary stage between the DA and RT stage is considered to mitigate the impact of forecast errors, which uses updated forecasts made closer to the RT dispatch time. The underlying ideology is that increased forecast accuracy is observed as the forecast interval gets closer to the actual dispatch time \cite{da_mg_sch_op2}. A pictorial representation of the proposed HMTS framework is shown in Fig. \ref{fig:layout}.
\subsection{Nomenclature}
The nodes belonging to the CMG are denoted by the set $\mathcal{N}_1$. The distribution system, external to CMG, is divided into smaller node groups (NGs) with nodes in each NG contained in the set $\mathcal{N}_{n > 1}$. The set $\mathcal{N}^{\text NG}=\{1,\dots, n\}$ contains the different NGs in the network. $\mathcal{N}_{n}^{\text {PV/ES/DG/CL/NCL}} \subseteq \mathcal{N}_{n}$ represents the set of nodes containing PV generators, energy storage units (ES), diesel generators (DG), critical loads (CLs), and non-critical loads (NCLs), respectively. $\mathcal{N}$ denotes the final set formed by aggregating the nodes belonging to the different node groups. The set $\mathcal{E} = \{(i,j): i \in \mathcal{N}, j \in \mathcal{N}, i \neq j\}$ denotes all the edges in the network and $\mathcal{E}_i$ is the set of all edges containing the node $i$. $s \in \Omega$ is the set of scenarios and $t \in \mathcal{T}$ is the set of hourly time intervals for the DA stage. Accordingly, $h \in \mathcal{H}_{t}$ and $k \in \mathcal{K}_{t,h}$ represent the sub-hourly intervals for NRT and RT stage corresponding to the $t^{\text {th}}$ hour and $h^{\text {th}}$ sub-hourly interval, respectively. $\Delta t/ \Delta h/ \Delta k$ are the time interval length for each stage. The parameters and variables in boldface indicate $3 \times 1$ dimensional vectors to represent values for the three network phases. $\langle \boldsymbol{\cdot}, \boldsymbol{\cdot} \rangle$ represent the inner product of two vectors. The vector valued variables with the tilde ($\tilde{\boldsymbol{\cdot}}$) represent the circularly shifted array wherein the elements are shifted by one position. The parameters containing the circumflex ($\hat{\boldsymbol{\cdot} }$) take the values obtained from the results of the immediate previous stage. Unless  explicitly specified, ($\underline{\boldsymbol{\cdot}},\overline{\boldsymbol{\cdot}}$) represents the minimum and maximum ratings, respectively.
\subsection{Stage 1: Day-Ahead Scheduling}
 A stochastic optimization-based formulation incorporating uncertainty in load and PV generation is solved for a period of the projected outage duration. This problem can be solved in a receding horizon fashion after every hour, or when there is any change in the outage duration, or when a significantly high forecasting error is observed. In such cases, this stage can be addressed as variable time frame scheduling. The objective function (\ref{eq:da_opt_obj}) aims at maximizing the total served load by prioritizing the critical loads. $\pi_s$ is the scenario probability, and $\omega_i$ is the load-based priority weight:  
\begin{equation}
\begin{aligned}
\vspace{-0.2cm}
\max_{P_{}^{\text {D}}} &\quad\sum\limits_{t \in \mathcal{T}} \sum\limits_{s \in \Omega} {{\pi _s}} {{\sum\limits_{n \in \mathcal{N}^{\text NG}, i \in \mathcal{N}_n} {{\omega _i}P_{i,t,s}^\text{D}}}}  \label{eq:da_opt_obj}.
\vspace{-0.5cm}
\end{aligned}
\end{equation}
The constraints are incorporated for all $n \in \mathcal{N}^{\text NG}$, $i \in \mathcal{N}_n$, $t \in \mathcal{T}$, and $s \in \Omega$, unless explicitly stated. Equations (\ref{eq:da_opt_p})--(\ref{eq:da_opt_q}) list the power balance constraints. This stage does not account for the detailed OPF constraints and instead emphasizes on integrating stochasticity to maintain low computation complexity while simultaneously addressing the forecasting error. Hence, a single-phase equivalent model of the CMG is considered by aggregating the values of variables and parameters for different phases at every node: 
\begin{subequations}
\begin{gather}
 \sum\limits_{n \in \mathcal{N}^{\text {NG}}, i \in \mathcal{N}^{\text {PV}}_n \cup {\mathcal{N}^{\text {DG}}_n} \cup {\mathcal{N}^{\text {DG}}_n}}{P_{i,t,s}^{\text {G}}}  = \sum\limits_{n \in \mathcal{N}^{\text {NG}}, i \in \mathcal{N}_n} {P_{i,t,s}^{\text {D}}},  \label{eq:da_opt_p}\\
\sum\limits_{n \in \mathcal{N}^{\text {NG}}, i \in \mathcal{N}^{\text {PV}}_n \cup {\mathcal{N}^{\text {DG}}_n} \cup {\mathcal{N}^{\text {DG}}_n}} {Q_{i,t,s}^{\text {G}}}  = \sum\limits_{n \in \mathcal{N}^{\text {NG}}, i \in \mathcal{N}_n} {Q_{i,t,s}^{\text {D}}}.  \label{eq:da_opt_q}
%
%
%
\end{gather}
\end{subequations}
Equations (\ref{eq:da_opt_pvds1})--(\ref{eq:da_opt_pvds3}) represent the real, reactive, and apparent power limits of the PV generators for all $i \in \mathcal{N}^{\text {PV}}_n$. $\overline P _{i,t,s}$ takes the DA forecast value. $\theta_{n,t}$ is a binary variable indicating the connectivity status of NG $n$: 
\begin{subequations}
\begin{gather}
0 \le P_{i,t,s}^{\text {PV}} \le \theta_{n,t}\overline P _{i,t,s}^{\text {PV,DA}},  \label{eq:da_opt_pvds1}\\
%
0 \le Q_{i,t,s}^{\text {PV}} \le \theta_{n,t}\overline Q _{i,t,s}^{\text {PV,DA}},  \label{eq:da_opt_pvds2}\\
[{(P_{i,t,s}^{\text {PV}})^2} + {(Q_{i,t,s}^{\text {PV}})^2}] \le {(\overline S _i^{\text {PV,DA}})^2}. \label{eq:da_opt_pvds3} 
\end{gather}
\end{subequations}
Equations (\ref{eq:da_opt_esds1})--(\ref{eq:da_opt_esds5}) represent the ES real power, reactive power, apparent power, and inter-temporal SOC change constraints for all $i \in \mathcal{N}^{\text {ES}}_n$. $\gamma$ is the reserve factor that ensures the ES is not operating at its limits to allow space for reserves. $P_{i,t,s}^{\text {ES}} > 0$ indicates battery discharge, and vice versa:
\begin{subequations}
\begin{gather}
 - \theta_{n,t}\overline P _i^{\text { ES,DA}} \le \gamma P_{i,t,s}^{\text {ES}} \le \theta_{n,t} \overline P _i^{\text {ES},DA},  \label{eq:da_opt_esds1}\\
 0 \le \gamma Q_{i,t,s}^{\text {ES}} \le \theta_{n,t}\overline Q _{i}^{\text {ES},DA} , \label{eq:da_opt_esds2}\\
  [{(P_{i,t,s}^{\text {ES}})^2} + {(Q_{i,t,s}^{\text {ES}})^2}] \le {(\overline S _i^{\text {ES},DA})^2} , \label{eq:da_opt_esds3}\\
 SO{C_{i,t,s}^{\text {ES}}} = SO{C_{i,t - 1,s}^{\text {ES}}} - \frac{{P_{i,t,s}^{\text {ES}}}}{{E_i^{\text {ES},DA}}}\Delta t\ , \label{eq:da_opt_esds4}\\
 \underline {SOC} _i^{\text {ES}} \le SOC_{i,t,s}^{\text {ES}} \le \overline {SOC} _i^{\text {ES}}.   \label{eq:da_opt_esds5}
\end{gather}
\end{subequations}
Equations (\ref{eq:da_opt_dgds1})--(\ref{eq:da_opt_dgds5}) represent the DG real power, reactive power, apparent power, ramp up/down, and inter-temporal fuel consumption constraints for all $i \in \mathcal{N}^{\text {DG}}_n$. 
$\alpha_i$ (L/hr-kW) and $\beta_i$ (L/hr-kW) are the the fuel consumption coefficients \cite{dg_fuel_eq}:
\begin{subequations}
\begin{gather}
\theta_{n,t}\underline P _i^{\text { DG,DA}} \le \gamma P_{i,t}^{\text {DG}} \le \theta_{n,t} \overline P _i^{\text { DG,DA}}, \label{eq:da_opt_dgds1}\\ 
\theta_{n,t} \underline Q _i^{\text { DG,DA}} \le \gamma  Q_{i,t}^{\text {DG}} \le \theta_{n,t}\overline Q _i^{\text { DG,DA}}, \label{eq:da_opt_dgds2} \\
[{(P_{i,t}^{\text {DG}})^2} + {(Q_{i,t}^{\text {DG}})^2}] \le {(\overline S _i^{\text { DG,DA}})^2},  \label{eq:da_opt_dgds3}\\
P_i^{\text {DG,RD,DA}} \le P_{i,t}^{\text {DG}} - P_{i,t - 1}^{\text {DG}} \le P_i^{\text {DG,RU,DA}} , \label{eq:da_opt_dgds4}\\
F_{i,t}^{\text {DG}} = F_{i,t - 1}^{\text {DG}} - ({\alpha_i}P_{i,t}^{\text {DG}}+ {\beta_i}\theta_{n,t}\overline P _i^{\text {DG}})\Delta t .  \label{eq:da_opt_dgds5}
%
\end{gather}
\end{subequations}
Equations (\ref{eq:da_opt_ld6})--(\ref{eq:da_opt_ld8}) place limits on CL and NCL load. $\underline P_{i,t,s}^{\text {D}}$, $\underline Q_{i,t,s}^{\text {D}}$ are set to $0$ for $i \in \mathcal{N}^{\text {NCL}}_n$ and to a minimum must-supply value for $i \in \mathcal{N}^{\text {CL}}_n$. $\overline P_{i,t,s}^{\text {D}}$, $\overline Q_{i,t,s}^{\text {D}}$ take the DA forecast value:
\begin{subequations}
\begin{gather}
%
%
%
%
%
\theta_{n,t}\underline P_{i,t,s}^{\text {D}} \le P_{i,t,s}^{\text {D}} \le \theta_{n,t} \overline P _{i,t,s}^{\text {D}},   \label{eq:da_opt_ld6}\\
%
%
\theta_{n,t}\underline Q _{i,t,s}^{\text {D}} \le Q_{i,t,s}^{\text {D}} \le \theta_{n,t}\overline Q _{i,t,s}^{\text {D}}.   \label{eq:da_opt_ld8}
\end{gather}
\end{subequations}
Equation (\ref{eq:da_opt_ng1}) establishes a minimum service duration (MSD) of $\upsilon$ hours for which a NG $n \in \mathcal{N}^{NG} \backslash \{1\}$ must stay connected:  
\begin{equation}
\begin{aligned}
\sum\limits_{t' = t}^{t + (\upsilon  - 1)} {{\theta _{n,t'}}}  \ge \upsilon ({\theta _{n,t}} - {\theta _{n,t - 1}}). \label{eq:da_opt_ng1}
\end{aligned}
\end{equation}

\subsection{Stage 2: Near Real-Time Schedule Update}
 This OPF-based problem is solved on a hourly basis with sub-hourly intervals after the occurrence of outage. The output is the updated referral plan and the load connectivity status. The objective function (\ref{eq:obj_nrt}) is modeled to minimize the squared error between the following: forecasted load and supplied load; DG power output and the reference DA value for the $t^{\text {th}}$ hour; ES SOC and the reference DA SOC at the end of the $t^{\text {th}}$ hour:
\begin{subequations}
\begin{gather*}
\label{eq:obj_nrt}
{\min _{{\textbf{\tiny P}}, {\textbf{\tiny SOC}}}}{\sum\limits_{h \in \mathcal{H}_t} \sum\limits_{i \in \mathcal{N}} {\omega _i} [\langle {\bf{1}},  {\bf{P}}_{i,h}^{\text {D}} - {\bf{\overline{P}}}_{i,h}^{\text {D}}} \rangle]^2 + \sum\limits_{i \in \mathcal{N}^{\text{DG}}}[\hat{P}_{i,t}^{\text {DG}} - \langle {\bf{1}},{\bf{P}}_{i,h}^{\text {DG}}\rangle]^2 \\ + \sum\limits_{i \in \mathcal{N}^{\text{ES}}}[( \hat{SOC}_{i,t}^{\text {ES}}- \langle {\bf{1}},{\bf{SOC}}_{i,\vert \mathcal{H}_t\vert}^{\text {ES}} \rangle )\Delta h / E_i^{\text {ES,DA}}]^2. \tag{8} 
\end{gather*}
\end{subequations}
For every hour, the NG connectivity status ($\theta_{n,t}$) is taken from the DA solution and the CMG node set is appended with the nodes of the connected NGs, thus forming the larger node set $\mathcal{N}$. For any given hour $t$, the constraints are incorporated for all $i \in \mathcal{N}$, $(i,j) \in \mathcal{E}$, and $h \in \mathcal{H}_t$, unless explicitly stated. Equations (\ref{eq:nrt_opt_pf1})--(\ref{eq:nrt_opt_pf2}) represent the nodal power balance equations using the branch flow model. Equations (\ref{eq:nrt_opt_pf5})--(\ref{eq:nrt_opt_pf7}) impose limits on the maximum power flowing over a network line and unidirectionality of power flow over a line using the binary variable vector ${\boldsymbol{\rho}}_{ij,h}$.  Equations (\ref{eq:nrt_opt_pf8})--(\ref{eq:nrt_opt_pf10}) compute the node voltages and limit the nodal voltages within acceptable bounds. ${{\boldsymbol{\zeta }}_{ij,h}}$ is a slack variable added to avoid conflict with (\ref{eq:nrt_opt_pf9}) when any two adjoining nodes are disconnected for a given direction of power flow:
\begin{subequations}
\begin{gather}
 {\bf{P}}_{i,h}^{\text {G}}-{\bf{P}}_{i,h}^{\text {D}}{=}\sum\limits_{j:ij \in \mathcal{E}_{i}}{{{\bf{P}}_{ij,h}}{-}\sum\limits_{i:ji \in \mathcal{E}_{i}}{{{\bf{P}}_{ji,h}}}}, \label{eq:nrt_opt_pf1} \\
{\bf{Q}}_{i,h}^{\text {G}} - {\bf{Q}}_{i,h}^{\text {D}} {=} \sum\limits_{j:ij \in \mathcal{E}_{i}} {{{\bf{Q}}_{ij,h}} {-} \sum\limits_{i:ji \in \mathcal{E}_{i}} {{{\bf{Q}}_{ji,h}}} }, \label{eq:nrt_opt_pf2}\\
0 \le {{\bf{P}}_{ij,h}} \le {{\boldsymbol{\rho}} _{ij,h}}\overline {\bf{P}}, \label{eq:nrt_opt_pf5}\\
0 \le {{\bf{Q}}_{ij,h}} \le {{\boldsymbol{\rho}} _{ij,h}}\overline {\bf{Q}}, \label{eq:nrt_opt_pf6}\\
{{\boldsymbol{\rho}}_{ij,h}} + {{\boldsymbol{\rho}}_{ji,h}} \le 1, \label{eq:nrt_opt_pf7}\\
{{\bf{V}}_{i,h}} \approx {{\bf{V}}_{j,h}} - {{\bf{A}}_{ij}}{{\bf{P}}_{ij,h}} - {{\bf{B}}_{ij}}{{\bf{Q}}_{ij,h}} + {{\boldsymbol{\zeta }}_{ij,h}} ,\label{eq:nrt_opt_pf8} \\
\underline {\bf{V}}  \le {{\bf{V}}_{i,h}} \le \overline {\bf{V}} ,\label{eq:nrt_opt_pf9} \\
- (1 - {{\boldsymbol{\rho}} _{ij,h}})\overline {\bf{V}} \; \le {{\boldsymbol{\zeta }}_{ij,h}} \le (1 - {{\boldsymbol{\rho}} _{ij,h}})\overline {\bf{V}},   \label{eq:nrt_opt_pf10}
\end{gather}
\end{subequations}
where
\begin{align}
\nonumber
\small {A_{ij}} = \left[ {\begin{array}{*{20}{c}}
{ - 2r_{_{ij}}^{\text {aa}}}&{r_{_{ij}}^{\text {ab}} - \sqrt 3 x_{_{ij}}^{\text {ab}}}&{r_{_{ij}}^{\text {ac}} + \sqrt 3 x_{_{ij}}^{\text {ac}}}\\
{r_{_{ij}}^{\text {ba}} + \sqrt 3 x_{_{ij}}^{\text {ba}}}&{ - 2r_{_{ij}}^{\text {bb}}}&{r_{_{ij}}^{\text {bc}} - \sqrt 3 x_{_{ij}}^{\text {bc}}}\\
{r_{_{ij}}^{\text {ca}} - \sqrt 3 x_{_{ij}}^{\text {ca}}}&{r_{_{ij}}^{\text {cb}} + \sqrt 3 x_{_{ij}}^{\text {cb}}}&{ - 2r_{_{ij}}^{\text {cc}}}
\end{array}} \right]
\end{align}
and
\begin{align}
\nonumber
\small {B_{ij}} = \left[ {\begin{array}{*{20}{c}}
{ - 2x_{_{ij}}^{\text {aa}}}&{x_{_{ij}}^{\text {ab}} + \sqrt 3 r_{_{ij}}^{\text {ab}}}&{x_{_{ij}}^{\text {ac}} - \sqrt 3 r_{_{ij}}^{\text {ac}}}\\
{x_{_{ij}}^{\text {ba}} - \sqrt 3 r_{_{ij}}^{\text {ba}}}&{ - 2x_{_{ij}}^{\text {bb}}}&{x_{_{ij}}^{\text {bc}} + \sqrt 3 r_{_{ij}}^{\text {bc}}}\\
{x_{_{ij}}^{\text {ca}} + \sqrt 3 r_{_{ij}}^{\text {ca}}}&{x_{_{ij}}^{\text {cb}} - \sqrt 3 r_{_{ij}}^{\text {cb}}}&{ - 2x_{_{ij}}^{\text {cc}}}
\end{array}} \right].
\end{align}
Constraints (\ref{eq:da_opt_pvds1})--(\ref{eq:da_opt_pvds3}) for PV, (\ref{eq:da_opt_esds1})--(\ref{eq:da_opt_esds5}) for ES, and (\ref{eq:da_opt_dgds1})--(\ref{eq:da_opt_dgds4}) for DG can be incorporated from the DA stage by making the following changes: replace $t$, $\mathcal{T}$, and $\Delta t$ by $h$, $\mathcal{H}_{t}$, and $\Delta h$; remove the scenario index $s$; remove NG connectivity indicator $\theta_{n,t}$; replace the single-phase variables and parameters with their equivalent three-phase vectors. Additional constraints (\ref{eq:nrt_opt_dgds5})--(\ref{eq:nrt_opt_dgds7}) pertaining to the DG phase imbalance and fuel consumption are added to this stage:
\begin{subequations}
\begin{gather}
%
%
%
%
%
{\bf{P}}_{i,h}^{\text {DG}} - {\bf{\tilde{P}}}_{i,h}^{\text {DG}} = {{\boldsymbol{\delta }}_{i,h}^{\text {DG}}} ,  \label{eq:nrt_opt_dgds5}\\
- {\overline {\boldsymbol{\delta }} _i^{\text {DG,NRT}}} \le {{\boldsymbol{\delta }}_{i,h}^{\text {DG}}} \le {\overline {\boldsymbol{\delta }} _i^{\text {DG,NRT}}}  , \label{eq:nrt_opt_dgds6}\\
F_{i,h}^{\text {DG}} = F_{i,h - 1}^{\text {DG}} - ({\alpha_i}\langle {\bf{1}},{\bf{P}}_{i,h}^{\text {DG}}\rangle  + {\beta_i}\overline P _i^{\text {DG}})\Delta h.  \label{eq:nrt_opt_dgds7}
%
\end{gather}
\end{subequations}
Equations (\ref{eq:nrt_opt_ld1})--(\ref{eq:nrt_opt_ld2}) computes the node load by incorporating the load connectivity status decision variable $x_{i}$, which stays fixed for all $h \in \mathcal{H}_t$:
\begin{subequations}
\begin{gather}
{\bf{P}}_{i,h}^{\text {D}} = {x_{i}}\overline {\bf{P}} _{i,h}^{\text {D}}, \label{eq:nrt_opt_ld1}\\
{\bf{Q}}_{i,h}^{\text {D}} = {x_{i}}\overline {\bf{Q}} _{i,h}^{\text {D}}.  \label{eq:nrt_opt_ld2}
%
%
\end{gather}
\end{subequations}
To add MSD to every load node, a new set $\mathcal{N}^{\text {MSD}}$ is introduced, which contains the nodes that have been connected and must remain connected until the MSD is completed. Hence, the constraint $x_{i} = 1 \hspace{0.2cm} \forall i \in \mathcal{N}^{\text {MSD}}$ is added.

\subsection{Stage 3: Real-Time Dispatch}
This stage is solved using the RT load and renewable generation data by splitting each sub-hourly interval into equal time slots with smaller time granularity. The objective (\ref{eq:obj_rt}) is to minimize PV curtailment, which ensures that the excess PV generation is incentivized to charge the ES:
\begin{equation}
\begin{aligned}
\label{eq:obj_rt}
{\min _{PV}}{\sum\limits_{i \in \mathcal{N}^{\text {PV}}}} [ \langle {\bf{1}}, \overline{\bf{P}}_{i,k}^{\text {PV}}-{\bf{P}}_{i,k}^{\text {PV}} \rangle]^2.
\end{aligned}
\end{equation}
For every interval $k$ within the sub-hour $h$ of hour $t$, the load connectivity decision is obtained as a parameter from the NRT stage solution. For every constraint incorporated from the NRT stage, $h$, $\mathcal{H}_{t}$, and $x_i$ are replaced by $k$, $\mathcal{K}_{t,h}$, and $\hat{x}_i$, respectively.  
The OPF constraints (\ref{eq:nrt_opt_pf1})--(\ref{eq:nrt_opt_pf9}), load constraints (\ref{eq:nrt_opt_ld1})--(\ref{eq:nrt_opt_ld2}), and NRT equivalent PV generation constraints (\ref{eq:da_opt_pvds1})--(\ref{eq:da_opt_pvds3}), ES constraints (\ref{eq:da_opt_esds1})--(\ref{eq:da_opt_esds5}), and DG constraints (\ref{eq:da_opt_dgds3})--(\ref{eq:da_opt_dgds5}) are replicated from the NRT stage by incorporating the above listed changes. The ES devices operate in the droop-controlled grid-forming mode, while the DG units operate in PQ grid-following mode. Hence, the output of the DG units will be fixed at the values obtained from NRT stage solution as shown in (\ref{eq:rt_opt_dgds1})--(\ref{eq:rt_opt_dgds2}), while that of the ES units cannot be pre-specified until the RT dispatch problem is solved \cite{droop_op}: 
\begin{subequations}
\begin{gather}
{\bf{P}}_{i,k}^{\text {DG}} = {\bf{\hat P}}_{i,h}^{\text {DG}},    \label{eq:rt_opt_dgds1} \\
{\bf{Q}}_{i,k}^{\text {DG}} = {\bf{\hat Q}}_{i,h}^{\text {DG}}.    \label{eq:rt_opt_dgds2}
\end{gather}
\end{subequations} 
The RT model may encounter infeasibility due to the equality constraints for load. At such times, these constraints are relaxed using an equivalent inequality constraint with upper/lower bounds and the objective function is appended with additional term representing the maximization of the served load. 
\section{Results}
\begin{figure}[htb]
\vspace{-0.5cm}
  \centering
  \includegraphics[width = 0.96\linewidth ,keepaspectratio, trim={0cm 0cm 0cm 0cm},clip]{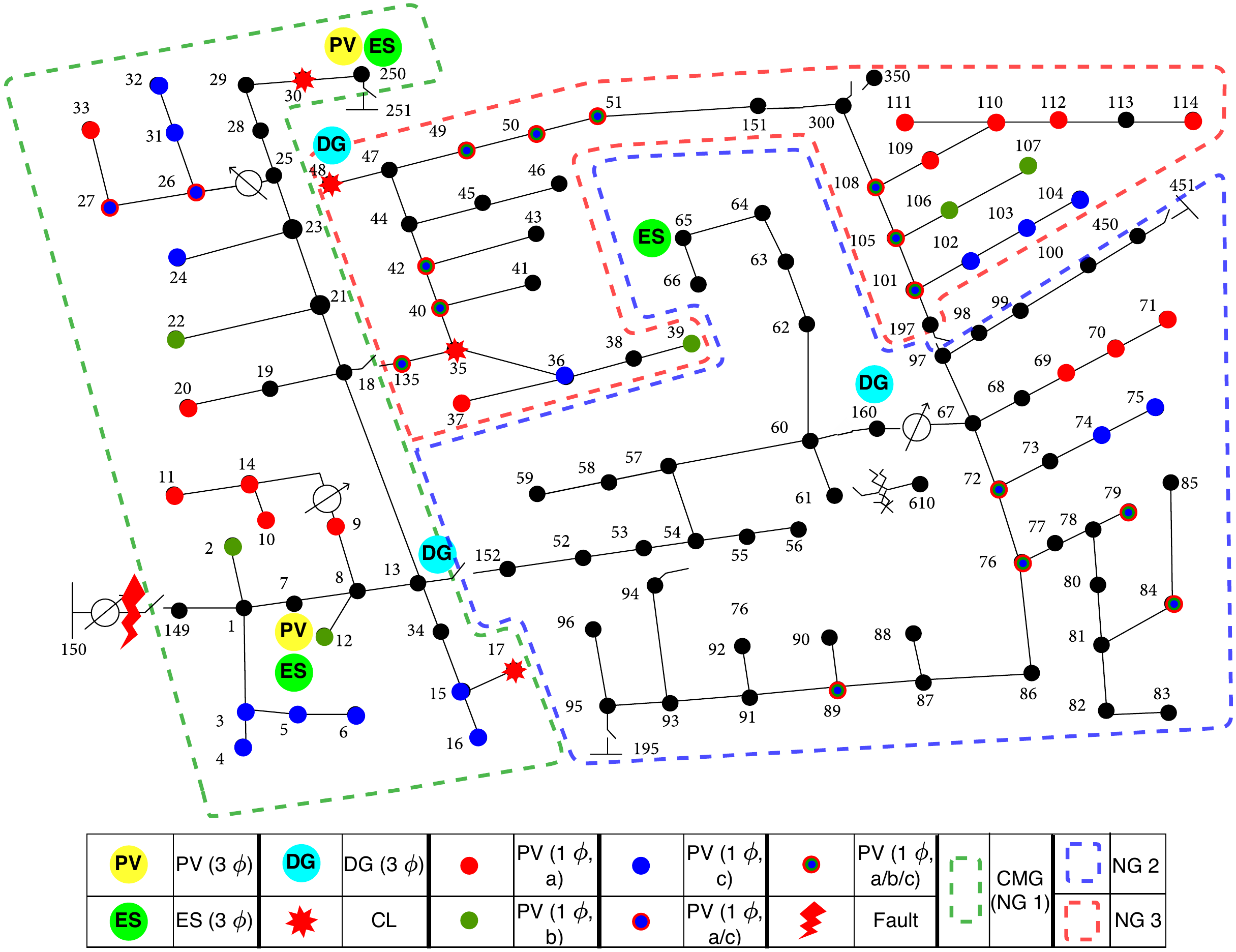}
  \vspace{-0.2cm}
  \caption{Modified IEEE 123 node system.}
  \label{fig:ieee123}
  \vspace{-0.5cm}
\end{figure}

\begin{figure} [htbp]
\vspace{-0.7cm}
\centering
  \subfloat[]{%
        \includegraphics[width=0.49\linewidth,keepaspectratio, trim={10cm 8cm 9.5cm 8cm},clip]{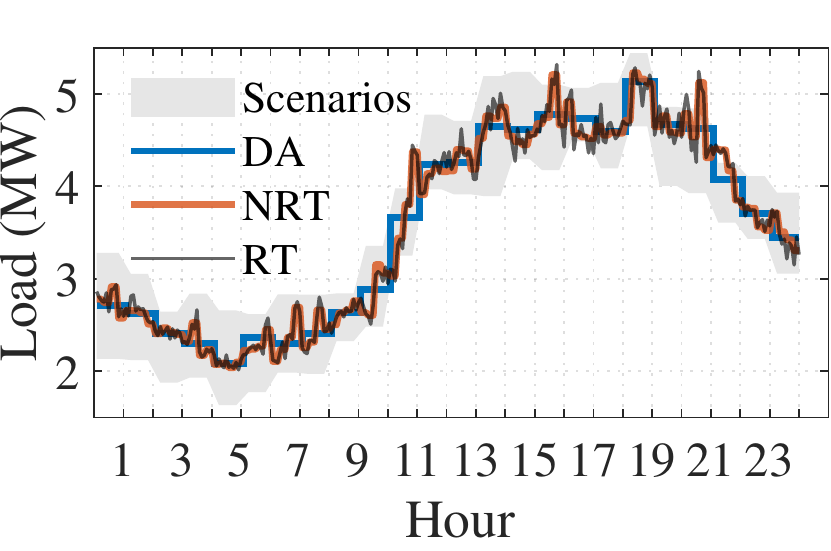}}
    \hfill
  \subfloat[]{%
        \includegraphics[width=0.49\linewidth,keepaspectratio, trim={10cm 8cm 9.5cm 8cm},clip]{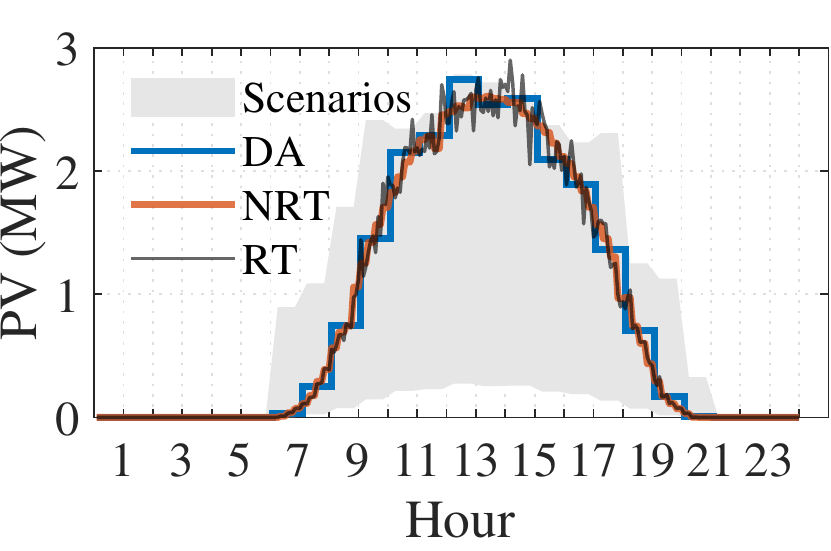}}
        \vspace{-0.2cm}
  \caption{Forecast profiles for base case: (a) Total load and (b) PV generation.}
  \label{fig:profiles} 
  \vspace{-0.25cm}
\end{figure}

The simulations have been performed on a modified IEEE $123$ bus system, as shown in Fig. \ref{fig:ieee123}. The NGs are formed on the basis of the location of pre-existing switches in the system. The distributed generation portfolio is listed in Table \ref{tab:dg_ratings}. The outage is assumed to occur at midnight and persists for a duration of $24$ hours. Figure \ref{fig:profiles} shows the base case forecasted load and PV generation profiles along with the bounds for the $40$ Monte-Carlo sampled DA scenarios. The base case NRT profiles are obtained from a utility in North Carolina, USA. The base case DA and RT profiles have been extrapolated from the NRT profile by adding a small random error sampled from a Gaussian distribution. The values for $\Delta t$, $\Delta h$, and $\Delta k$ are $1$ hour, $15$ minutes, and $5$ minutes, respectively. The load priority weight for CL and NCL is fixed at $2$ and $1$, respectively. The MSD for each NG and load node is $2$ hours. All ES units are assumed to be fully charged initially, and the SOC limits are $0.10$ and $0.95$. The reserve requirement is set to $20$\%. Each DG is equipped with $3000$ L of fuel.
The proposed MILP/MIQP formulation is solved in Python using CPLEX 12.10 solver on a PC with  Intel Core i9-9900K CPU @ 3.6GHz processor and 64 GB RAM.
\begin{table}[htbp]
\vspace{-0.35cm}
\centering
\caption{Distributed generation portfolio}
\vspace{-0.25cm}
\label{tab:dg_ratings}
\scalebox{0.92}{
\begin{tabular}{c|c|c}
\hline
Generator & Generator node  & Rating               \\ \hline \hline
DG${}^{*}$       & 13, 48, 160     & 600 kW, 600 kW,  50 kW \\ \hline
PV${}^{*}$       & 7, 250          & 750 kW, 750 kW         \\ \hline
ES${}^{*}$ & 7, 250, 65 & \begin{tabular}[c]{@{}c@{}}450 kW/900 kWh, 2.25 MW/4.5 MWh, \\ 100 kW/200 kWh\end{tabular} \\ \hline
BTM PV${}^{\#}$    & See Fig. \ref{fig:ieee123} & 3 to 15 kW               \\ \hline
\end{tabular}}
\begin{tablenotes}
\item[*] ${}^{*}$ = Three phase, ${}^{\#}$ = Single phase  
\end{tablenotes}
\end{table}

\begin{figure*} [tb]
\centering
  \subfloat[]{
    \includegraphics[width = 0.33\linewidth ,keepaspectratio, trim={4.75cm 10.62cm 4.5cm 10cm},clip]{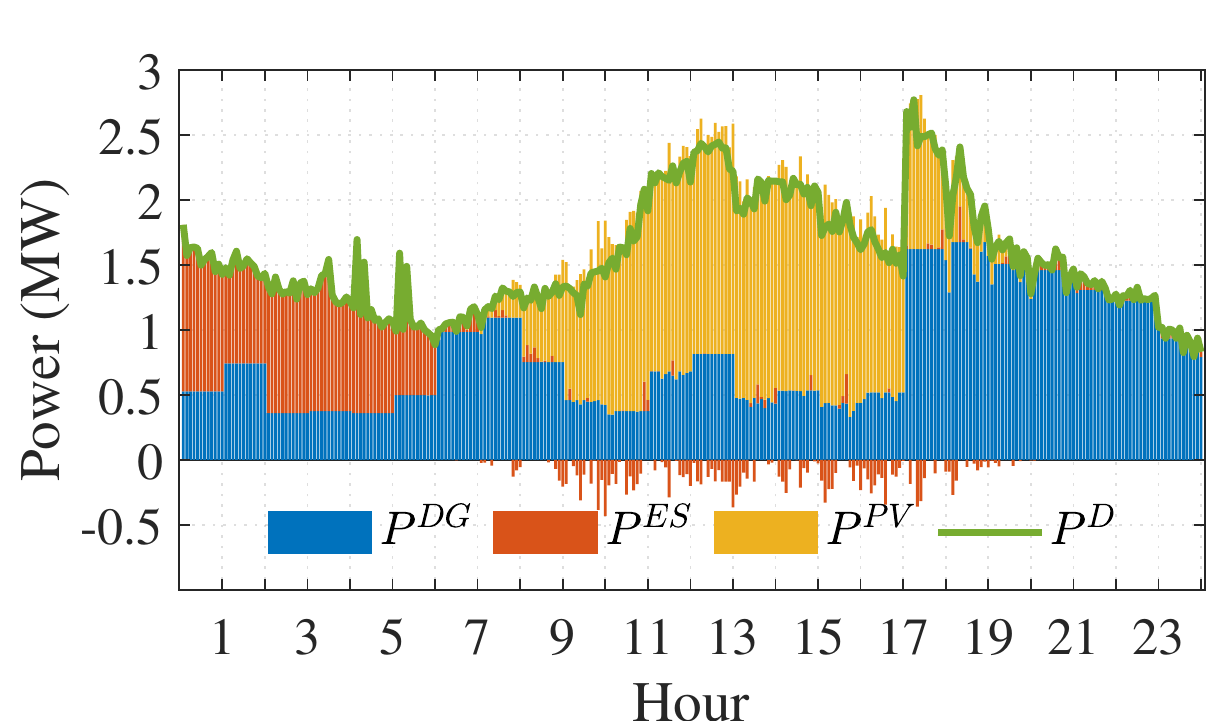}}
      \hfill
  \subfloat[]{%
        \includegraphics[width=0.33\linewidth,keepaspectratio, trim={4.75cm 11cm 4.5cm 10cm},clip]{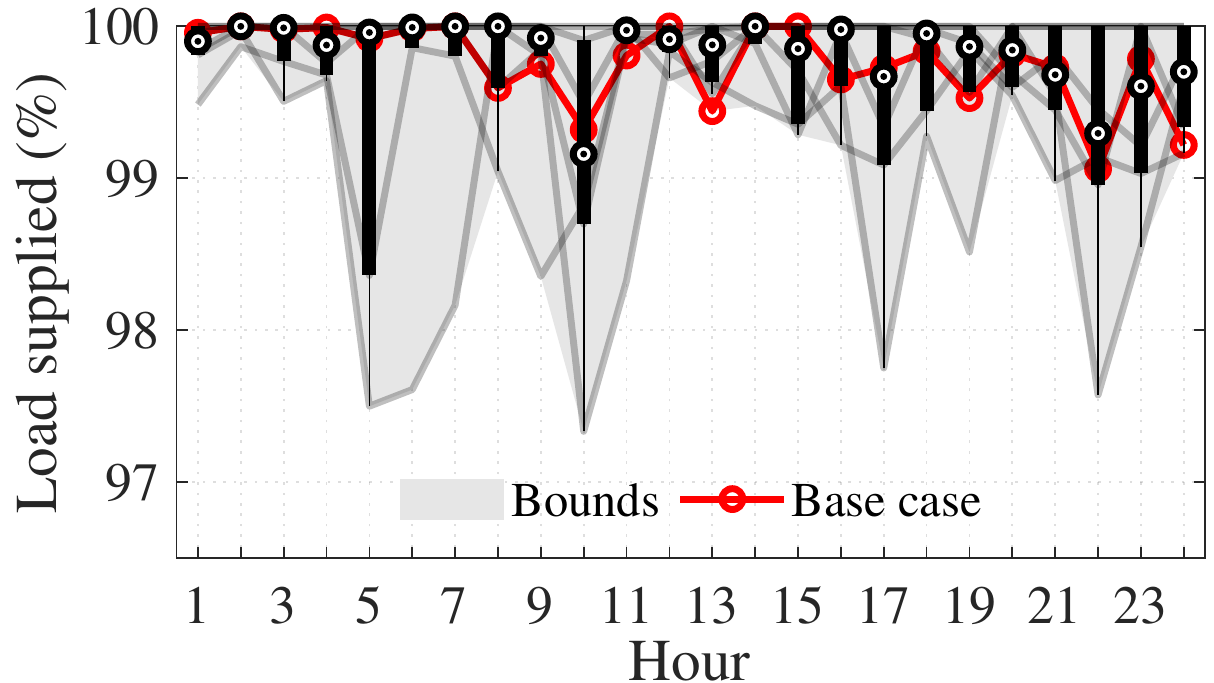}}
    \hfill
  \subfloat[]{%
        \includegraphics[width=0.33\linewidth,keepaspectratio, trim={4.75cm 11cm 4.5cm 10cm},clip]{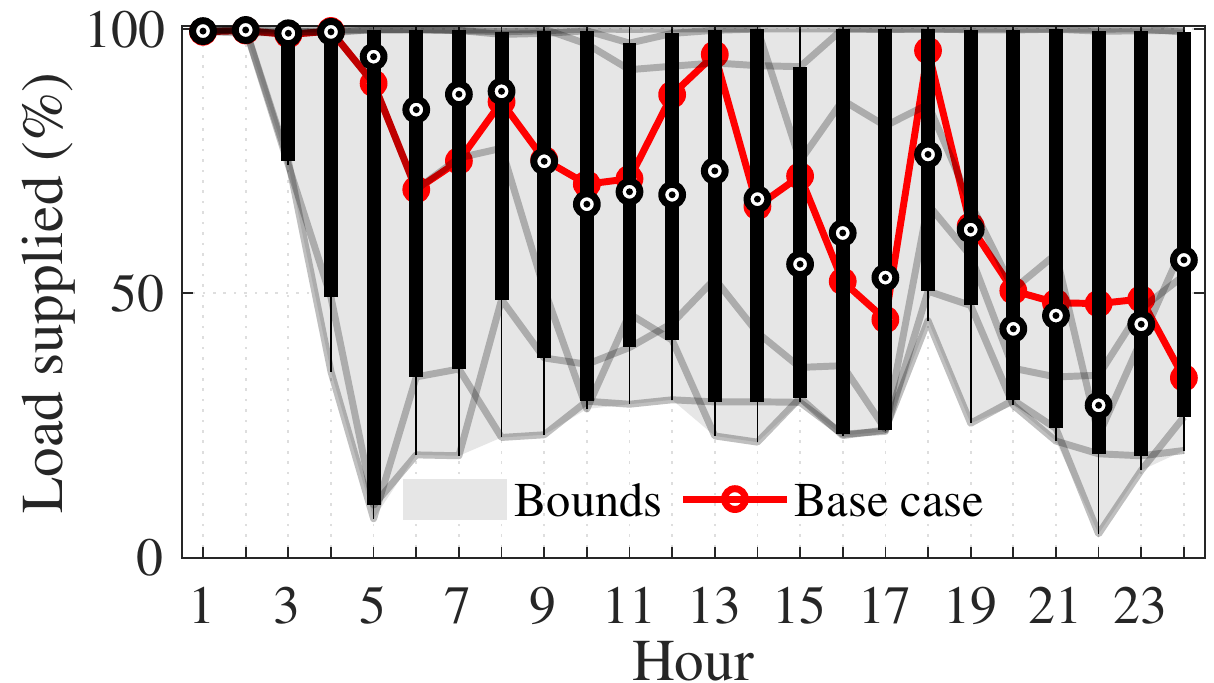}}
        \vspace{-0.20cm}
  \caption{HMTS simulation results: (a) Base case results, (b) Forecast error impact on CL demand, and (c) Forecast error impact on NCL demand.}
  \label{fig:results_all} 
\end{figure*}

The base case RT dispatch for the boundary extended CMG is shown in Fig. \ref{fig:results_all}(a). The CMG boundary was expanded to accommodate NG $3$ for the entire duration since it contains $2$ CL loads and higher generation capacity. We also observe that the resources are optimally allocated and ensure that they do not deplete until the outage has been resolved. The stochastic formulation ensures the DG output reference values, which are referred to by all subsequent stages, are carefully chosen. $77.41$\% of the total extended CMG load, and the entire CL demand was supplied in the RT dispatch. 

To numerically analyze the HMTS robustness to forecasting errors, we analyze its performance against various forecasting error cases between the three timescales. The six cases considered are: (A) NRT and RT PV forecast values are lower than DA forecast due to cloud cover; (B/C) NRT and RT load forecast is higher/lower than the DA forecast; (D/E) RT load forecast value is higher/lower than the NRT and DA forecast; (F) DA, NRT, and RT load forecast values follow a random pattern. The forecast error with regards to the base case is $\pm 50\%$. Ideally, as the forecast interval moves closer to the current time step, the prediction accuracy increases. Thus, cases D and E are extreme cases with a low probability of realization but have been added to study the model's robustness.  

We use two metrics to compare the HMTS framework performance. The first metric, shown in Fig. \ref{fig:results_all}(b) and Fig. \ref{fig:results_all}(c), indicates the percentage of the total CL and NCL demand that has been supplied for each hour for the base case and the forecast error cases A--F. The boxplot shows the variability in the total demand supplied for the six cases. The total CL demand supplied in all the cases deviates minimally from the base case value. However, the supplied NCL demand deviates significantly from the base case value to ensure that the supplied CL load is maximized. The second metric, shown in Table \ref{tab:metric2}, indicates the average duration for which the CL or NCL load is connected to the grid. For the proposed HMTS, the CL load is prioritized under all different cases by providing it connectivity to the grid for all time intervals. In contrast, the NCL load is frequently disconnected. This analysis helps conclude that under the different forecast error scenarios, the CMG can supply almost all CL demand and some portion of the NCL demand for the entire outage duration by sustainably allocating its limited resources. 
\begin{table}[]
\centering
\caption{Load connectivity duration}
\vspace{-0.25cm}
\label{tab:metric2}
\scalebox{0.96}{
\begin{tabular}{c|c|c|c|c|c|c|c}
\hline
Case                                                    &  Base & A  &  B  &  C  &  D  &  E  &  F  \\ \hline \hline
& \multicolumn{7}{c}{\bf{HMTS}}                                                             \\ \hline    
CL (\%)                                               & 100 & 100  & 100   & 100   & 100   & 100   & 100   \\ \hline
\begin{tabular}[c]{@{}c@{}}NCL (\%) \end{tabular} & 70.17 & 53.75 & 34.18 & 97.08 & 58.33 & 100 & 81.33 \\ \hline 
& \multicolumn{7}{c}{\bf{Two-stage}}                                                                             \\  \hline
CL (\%)                                               & 100 & 100   & 100   & 100   & 100   & 100   & 100   \\ \hline
\begin{tabular}[c]{@{}c@{}}NCL (\%)\end{tabular} & 59.58 & 53.33 & 45.83 & 85.23 & 45.83 & 85.23 & 68.75 \\ \hline
\end{tabular}}
\vspace{-0.5cm}
\end{table}

We then perform a comparative analysis of the HMTS approach and the traditional two-stage approach of DA scheduling followed by RT dispatch. For the two-stage model, the HMTS DA formulation is combined with the load connectivity status selection problem. No change is made to the RT formulation. Due to the omission of the NRT stage, cases D and E are the same as cases B and C. From Table \ref{tab:metric2}, we observe that the CL connectivity duration is the same as that of the HMTS framework. The NCL connectivity duration varies and is lesser than the HMTS approach. However, Table \ref{tab:metric1} shows that the two-stage approach fails to meet the entire CL demand. 

Due to the mismatch between DA forecasts and RT realization, the DA load commitment decisions are not per the DA allocated resources to meet the RT demand, resulting in RT model infeasibility due to the load equality constraints. Hence, these constraints are relaxed if an infeasibility is encountered. Due to the relaxations, the total load supplied is below $100\%$ even when the load is connected to the grid at all times. In islanded mode operation, the relaxed solution can be implemented if a fast-acting demand response mechanism is present or loads are abruptly disconnected without satisfying the MSD requirements. Table \ref{tab:infeas} shows that the HMTS model requires very few relaxations compared to the two-stage model due to the buffer provided by the intermediary NRT stage. 

\begin{figure} [htbp]
\vspace{-0.9cm}
\centering
  \subfloat[]{%
        \includegraphics[width=0.49\linewidth,keepaspectratio, trim={9cm 8.4cm 10.75cm 8cm},clip]{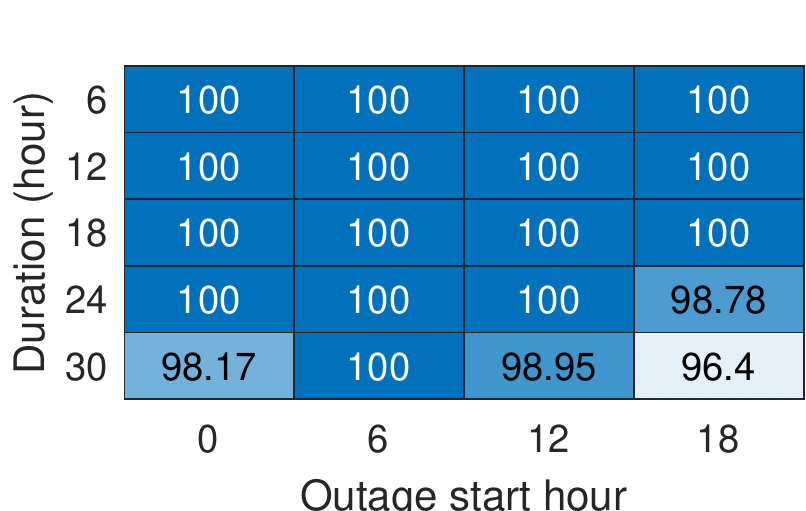}}
    \hfill
  \subfloat[]{%
        \includegraphics[width=0.49\linewidth,keepaspectratio, trim={9cm 8.4cm 10.75cm 8cm},clip]{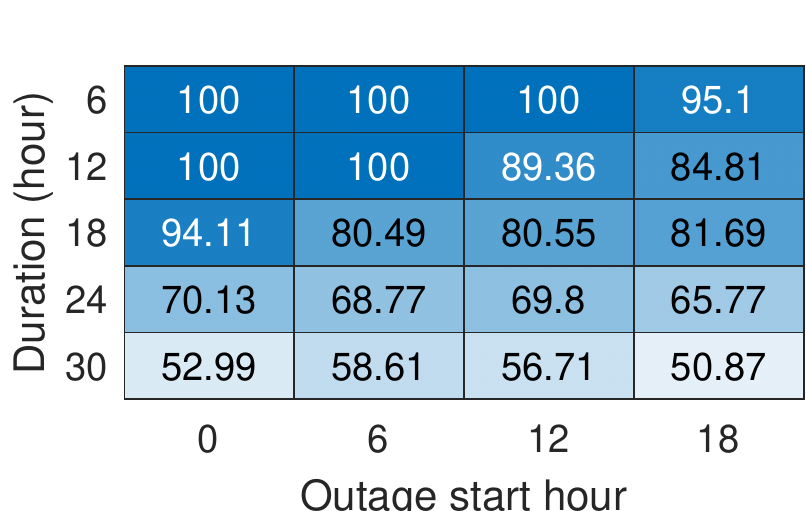}}
          \vspace{-0.15cm}
  \caption{Percent load supplied in outage scenarios: (a) CL and (b) NCL.}
  \label{fig:multitime}
  \vspace{-0.5cm}
\end{figure}

Next, the HMTS performance for multiple outage scenarios with different start times and duration is shown in Fig. \ref{fig:multitime}. For all but one scenario, the total CL demand supplied exceeds $98\%$, and at least $50\%$ of the NCL demand has been supplied. The median CL and NCL load supplied is $100\%$ and $81.12\%$, respectively. Minimum CL load supplied is $85.4\%$ with a median value of $97.27\%$ and minimum NCL load supplied is $28.92\%$ with a median value of $64.18\%$ for the two-stage approach. Overall, we conclude that the proposed HMTS approach serves almost the entire CL demand and a higher NCL demand than the two-stage approach under all different forecast error cases and outage scenarios without significantly relying on relaxed solutions that violate dispatch constraints.
\begin{table}[]
\centering
\vspace{-0.5cm}
\caption{Percent load supplied}
\vspace{-0.25cm}
\label{tab:metric1}
\scalebox{0.95}{
\begin{tabular}{c|c|c|c|c|c|c|c}
\hline
Case                                                    &  Base & A  &  B  &  C  &  D  &  E  &  F  \\ \hline \hline
& \multicolumn{7}{c}{\bf{HMTS}}                                                                                               \\  \hline
CL (\%)                                              & 100 & 99.80  & 99.73   & 100   & 98.26   & 100   & 99.83   \\ \hline
\begin{tabular}[c]{@{}c@{}}NCL (\%) \end{tabular} & 69.58 & 49.61 & 33.72 & 98.13 & 37.49 & 97.71 & 80.58 \\ \hline 
& \multicolumn{7}{c}{\bf{Two-stage}}                                                                             \\  \hline
CL (\%)                                               & 95.36 & 91.17   & 79.21   & 100   & 79.21   & 100   & 98.32   \\ \hline
\begin{tabular}[c]{@{}c@{}}NCL (\%)\end{tabular} & 43.71 & 39.61 & 28.45 & 52.91 & 28.45 & 52.91 & 45.81 \\ \hline
\end{tabular}}
\vspace{-0cm}
\end{table}
\begin{table}[]
\vspace{-0.4cm}
\centering
\caption{Percentage of RT intervals with relaxed solutions}
\vspace{-0.25cm}
\label{tab:infeas}
\scalebox{0.92}{
\begin{tabular}{c|c|c|c|c|c|c|c}
\hline
Approach       & Base                        &  A  &  B  &  C  &  D  &  E  &  F                      \\ \hline \hline
HMTS (\%)      & \multicolumn{1}{c|}{3.82} & 4.51  & 6.25  & 2.08  & 67.71 & 10.07 & \multicolumn{1}{c}{5.98} \\ \hline
$2$-stage (\%) & 37.15                     & 61.43 & 75.00 & 70.13 & 75.00 & 70.13 & 57.63                     \\ \hline
\end{tabular}}
\vspace{-0.5cm}
\end{table}
\vspace{-0cm}
\section{Conclusion}
This paper proposes an HMTS framework for proactive and resilient scheduling and dispatch of islanded CMGs during emergency conditions. A buffer stage between the DA scheduling and the RT dispatch is added to update the DA schedule using newly obtained forecasts closer to the actual RT dispatch time. The different CMG scheduling and dispatch constraints are implemented at different hierarchical stages based on the relevance of a particular constraint to a specific timescale, which minimizes the computational burden at each stage. Our results show that with different outage duration and combinations of forecast errors, the NRT stage provides an updated feasible schedule as a reference for the RT dispatch, which ensures the RT dispatch can securely meet the load demand. Our future work will involve more comprehensive criteria for making decisions on CMG boundary expansion, verification on real-world systems, and hardware-in-loop validation of the proposed HMTS.   
\vspace{-0.2cm}
\section*{Acknowledgement}
The authors thank PJ Rhem with ElectriCities, Paul Darden, Steven Hamlett, and Daniel Gillen with Wilson Energy for their inputs, suggestions, and technical guidance.
\bibliographystyle{IEEEtran}
\vspace{-0.20cm}
\bibliography{references}

\end{document}